# Near-field radiative heat transfer between rough surfaces modeled using effective media with gradient distribution of dielectric function

D. Y. Xu[a], A. Bilal[a], J. M. Zhao[a,c,*], L. H. Liu[a,b], Z. M. Zhang[d]

[a]*School of Energy Science and Engineering, Harbin Institute of Technology, Harbin 150001, China*
[b]*School of Energy and Power Engineering, Shandong University, Qingdao 266237, China*
[c]*Key Laboratory of Aerospace Thermophysics, Ministry of Industry and Information Technology, Harbin 150001, China*
[d]*G. W. Woodruff School of Mechanical Engineering, Georgia Institute of Technology, Atlanta, Georgia 30332, USA*

## Abstract

Near-field radiative heat transfer (NFRHT) between rough surfaces, due to its widespread presence in engineering practice of near-field energy utilization, requires indepth studies, especially from the perspective of physical mechanism. In this paper, an effective multilayer model is built to approach the NFRHT between random rough surfaces of silicon carbide (SiC). Using the effective medium theory (EMT), the effective dielectric function of each layer is obtained, which forms a gradient distribution of dielectric function along the depth of the medium. The influence of the effective dielectric function on surface phonon polaritons (SPhPs) is analyzed, showing that the effective layers with small filling fraction of SiC feature lower SPhP resonance frequencies than SiC bulk. The coupling of SPhPs from the gradient distribution of dielectric function produces new surface modes that dominates the NFRHT. Investigation on the effect of root mean square height (RMS height, $\sigma$) reveals that the peaks of local density of states (LDOS) and spectral heat flux are red-shifted as $\sigma$ increases, while the spectral heat flux below the peak frequency gets larger. This can be attributed to the coupling of SPhPs inside the rough layer. We also found the total net heat flux between rough surfaces separated by an average distance exceeds that between smooth plates and increases with increasing $\sigma$, which offer a new way to enhance NFRHT. This work provides a reference for the simulation and understanding of the NFRHT between rough surfaces.

---

[*] Corresponding author.
  *Email address*: jmzhao@hit.edu.cn (J. M. Zhao)

**Nomenclature**

| | |
|---|---|
| *dep* | depth of a point inside the medium, m |
| *N* | number of points |
| *rL* | total length of the rough surface, m |
| *x* | *x* coordinate of a point, m |
| *z* | *z* coordinate of a point, m |
| *P* ($\eta$) | probability of *z*=$\eta$ |
| *C* | autocovariance function, m$^2$ |
| *n* | number of effective layers |
| *t* | thickness of effective layers, m |
| *f* | filling fraction of SiC |
| *T* | temperature, K |
| *d* | vacuum gap distance, m |
| $d_{\mathrm{rms}}$ | defined in Section 2.3 |
| $d_{\mathrm{top}}$ | defined in Section 2.3 |
| $k_{\mathrm{B}}$ | Boltzmann constant, 1.381×10$^{-23}$ J/K |
| $q_{\mathrm{net}}$ | total net heat flux, W/m$^2$ |
| $q_{\omega}$ | spectral heat flux, Wm$^{-2}$s |
| *s* | exchange function |
| $k_0$ | magnitude of wavevector in vacuum, m$^{-1}$ |
| *c* | light speed in vacuum, 2.998×10$^8$ m/s |
| *R* | effective reflection coefficient |
| *D* | local density of states, LDOS |
| *p* | defined in section 3.2 |
| $h_{\mathrm{t}}$ | defined in section 3.2 |

*Greek Symbols*

| | |
|---|---|
| $\sigma$ | root mean square height, m |
| $\tau$ | correlation length, m |
| $\varepsilon$ | dielectric function |
| $\omega$ | angular frequency, rad/s |
| $\Gamma$ | damping factor, s$^{-1}$ |
| $\Theta$ | mean energy of a Planck oscillator, J |
| $\beta$ | parallel wavevector component, m$^{-1}$ |
| $\gamma$ | vertical wavevector component, m$^{-1}$ |
| $\hbar$ | Planck constant divided by 2$\pi$, 1.055×10$^{34}$ Js |

| $\rho$ | effective reflectivity |
|---|---|
| $\kappa$ | defined in section 3.2 |
| $\zeta$ | energy transmission coefficient |

*Subscripts*

| 0 | vacuum |
|---|---|
| 1 | medium 1 |
| 2 | medium 2 |
| vac | vacuum |
| eff | effective layer |
| res | SPhPs resonance |
| pea | peak of spectral heat flux |
| prop | propagating waves |
| even | evanescent waves |

*Superscripts*

| p | p polarization or TM waves |
|---|---|
| s | s polarization or TE waves |

# 1 Introduction

Radiative heat flux can exceed the limit of blackbody radiation when the separation distance becomes comparable with or shorter than the characteristic wavelength given by Wien's displacement law [1-9]. The main reason for the enhancement of radiative heat transfer in near-field regime is the excitation of surface waves, i.e., surface plasmon polaritons (SPPs) or surface phonon polaritons (SPhPs), and their photon tuning [10, 11]. Owing to its promising applications on energy conversion and management [9, 12-15], numerous studies have been conducted on the near-field radiative heat transfer (NFRHT), such as, the NFRHT between hyperbolic metamaterials [16-19], films [20-22] and phase-change materials [23-28], as well as the effect of graphene and other two-dimensional materials [29-33]. In these studies, the surfaces of plates are all assumed to be ideally smooth, which, however, cannot be satisfied in engineering application. Therefore, it is worthwhile to investigate the effect of rough surfaces on NFRHT.

Due to the geometric complexity of the surface and the difficulty in modeling, only few references discussed the NFHRT between rough surfaces. Biehs and Greffet [34] presented a second-order perturbation theory to determine the NFRHT between two semi-infinite media with rough surfaces, finding that the heat flux between two SiC slabs becomes larger when roughness is considered. They also showed the proximity approximation (PA), which was employed by Persson *et al.* [35] to consider roughness effects on NFRHT, is valid for the gap distance much smaller than the correlation length of the rough surface. Krüger *et al*. [36] studied the distance dependence of the interaction between curved rough surfaces using PA. They predicted that the NFRHT between a rough sphere and a plate approaches a constant with decreasing separation distance. Chen and Xuan [37, 38] discussed the influence of surface roughness on the NFRHT between two plates of Drude material using the finite difference time-domain (FDTD) method based on Wiener Chaos Expansion (WCE). Their results manifest that the NFRHT can be either suppressed or enhanced through adjusting the RMS height and the correlation length. Also by FDTD, Didari and Mengüç [39] studied the near-field thermal emission between corrugated surfaces. They found different degrees of enhancement compared to smooth surface, depending on the shapes and sizes of the surface. These researches provide valuable inspiration for the NFRHT between rough surfaces. However, to understand the effect of roughness physically and to reveal the mechanism of surface polaritons, the key role in NFRHT, theoretical approximation models still need to be established.

In this paper, we adopt a new method to approach the NFRHT between two media with rough surfaces made of a silicon carbide (SiC), a typical material supporting SPhPs. Modeling it using an effective multilayer medium with gradient distribution of dielectric function, we analyze the SPhPs resonances of these effective layers. The local density of states (LDOS) and spectral heat flux for rough surfaces are investigated to reveal the effect of root mean square height (RMS height) on NFRHT. The evolution of energy transmission coefficient is also explored to reveal the SPhPs' interaction mechanism and how the new surface modes emerges. In addition, we also verify the feasibility of our model when used for studying the effect of correlation length.

## 2 Theoretical model

The profile of the rough surface of a medium is shown in **Fig**. **1** (a) and a schematic of its cross section on *x-z* plane is shown in **Fig**. **1** (b). The dielectric function or refraction index varies along *z*-axis due to the variance of the material composition. To describe this feature more clearly, we divide the rough layer of rough surface into several layers as shown in **Fig**. **1** (b). Each layer is composed of two parts: SiC and vacuum. Effective medium theory (EMT) is adopted to obtain the effective dielectric function using the filling fraction of each layer. Here, we define *dep* as the vertical distance between an interested location of the medium and the highest point of the rough surface, the same below. As *dep* increases, the filling fraction of SiC grows and becomes unity when it comes to the interior of the medium. Correspondingly, the effective dielectric function varies from 1 (the value of vacuum) to the value of SiC. In this way, an effective multilayer model with gradient distribution of dielectric function is established, which is used to approach the NFRHT between rough surfaces in this work.

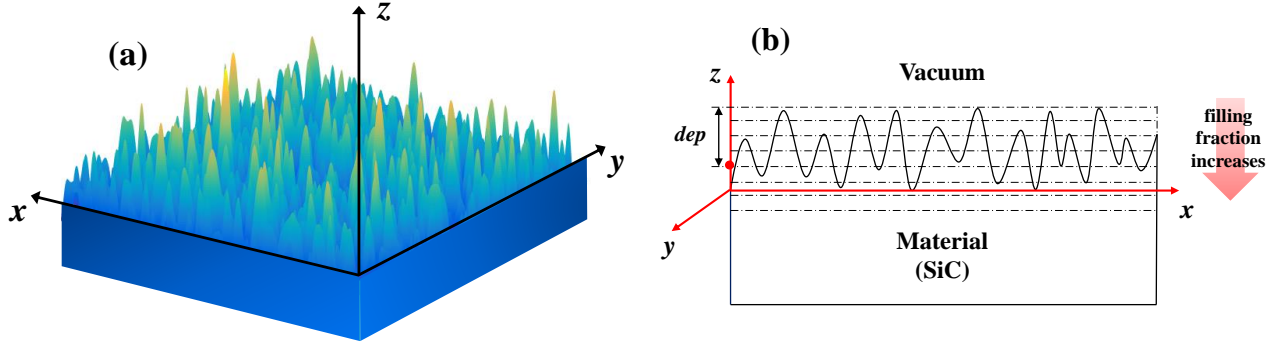

**Fig. 1** Profile of a medium with rough surface (a) and a schematic of its cross section on *x-z* plane (b) forming the gradient distribution of dielectric function. The rough layer is divided into several layers, with each layer featuring a filling fraction of SiC, which increases along *z*-axis. The definition of *dep* is also shown, which is the vertical distance between an interested location of the medium and the highest point of the rough surface.

## 2.1 Mathematical description of rough surface

Gaussian distribution is adopted to generate the random rough surface in this paper. First, *N* points are evenly spaced over the total length of the rough surface, which is represented by *rL* in the following. The location of each point is represented by (*x*, *z*), where *z* is the height of the point at *x*. Due to Gaussian distribution function, the probability of $z = \eta$, which is referred to as $P(\eta)$, is expressed as [40]

$$P(\eta) = \frac{1}{\sigma\sqrt{2\pi}} e^{-\eta^2/2\sigma^2} \quad (1)$$

where, $\sigma = \sqrt{\langle \eta^2 \rangle}$ is the root mean square (RMS) height. Here, $\langle \ \rangle$ means the ensemble average of the *N* points. $\sigma$ is a parameter describing the vertical irregularities of a rough surface, i.e., the roughness.

The autocovariance function describes the variance of these points laterally along the surface (i.e., the crowdedness of the hills and valleys). Here we define the autocovariance function using a Gaussian function [40]

$$C(\tau) = \langle \eta(x_1)\eta(x_2) \rangle = \sigma^2 \exp(-\frac{|x_1 - x_2|^2}{\tau^2}) \quad (2)$$

where $x_1$ and $x_2$ are the *x* coordinates of two points along the surface and $\tau$ is correlation length, the typical distance between two similar features (e.g., hills or valleys).

For modelling and simulative purpose, rough surfaces with Gaussian statics can be generated using a method outlined by Garcia and Stoll [41] where an uncorrelated distribution of surface points generated by a random number generator (i.e., white noise) is convolved with a Gaussian filter to achieve correlation. This convolution is most efficiently performed using the discrete Fast Fourier Transform (FFT) algorithm, which in MATLAB is based on the FFTW library. It's necessary to point out that when generating a rough surface, we take the opposite of $\eta$ with a negative value to ensure that the lowest point of the rough layer is at $z = 0$. Obviously, this operation does not change the value of $\sigma$ which only depends on the value of $\eta^2$.

**Fig. 2**(a) shows the random rough surface generated according to the parameters of $N = 1000$, $rL = 50$ μm, $\tau = 0.1$ μm, $\sigma = 10$ nm. As can be seen, the minimum value of $z$ is 0 while the maximum value reaches about 40 nm. However, points with $z$ coordinates over $3\sigma$ (here, 30 nm) are very rare, accounting for only 0.4% of the total number of points. What's more, if the points on both sides get extremely high, even touching each other, the interaction of these points through radiation and even conduction will dominate the heat transfer. In that case, our model which treats both surfaces as effective multilayer media will lose its validaty. Actually, for real random surfaces with micro roughness, which is usually made from some polishing process, hence it is about impossible for some extremly high (singular) point to appear. Therefore, hereafter, we neglect the points with z coordinates over $3\sigma$.

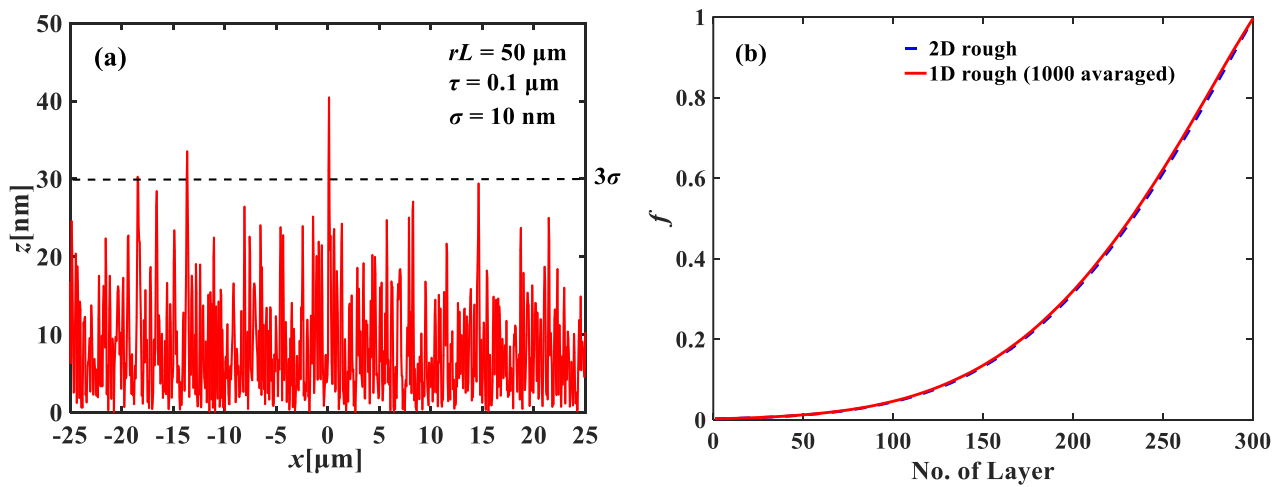

**Fig. 2** (a) Random rough surface profile generated through the method above using parameters: $N = 1000$, $rL = 50$ μm, $\tau = 0.1$ μm, $\sigma = 10$ nm. The dashed line represents the reference location of $z = 3\sigma$. (b) Filling fraction of SiC as a function of the serial number of layers. The total number of layers is 20 and the thickness of each layer is 1.6 nm. The serial number of layers increases with increasing *dep*. (i.e., the 1st layer is at the top and the 20th layer is at the bottom).

If the rough layer of rough surface is divided into 300 layers and the thickness of each layer is 0.1 nm, we can get the filling fraction of SiC ($f$, the same below) as a function of the serial number of layers shown in **Fig. 2**(b) with red solid line. Here we take the average of 1000 random rough surfaces and the reason of this operation will be analyzed in the following discussion. One should notice that the serial number of layers increases as *dep* increases. (i.e., the 1$^{st}$ layer is at the top and the 20$^{th}$ layer is at the bottom). As shown in **Fig. 2**(b), the filling fraction of SiC increases from 0 (for the 1$^{st}$ layer) to 1 (for the 20$^{th}$ layer). As a result, the dielectric function gradually changes from the value of vacuum to the value of SiC, thus forming a gradient distribution as will be discussed in detail later.

Above discussion is about 1D random rough surfaces. For 2D rough random surfaces whose $\tau$ are the same in *x* and *y* axes (in this paper, only this kind of rough surface is considered), due to the isotropy in *x-y* plane, the profiles on all cross sections that parallel to *z*-axis show no essential differences and can be represented by 1D rough surface profile. As a result, if averaging substantially the $f$ obtained from 1D algorithm, the $f$ of 2D rough surface can be immediately obtained. Also shown in **Fig. 2**(b) with blue dashed line is the exact $f$ calculated using the 2D algorithm for the same parameters as **Fig. 2** (a). There is nearly no difference between this result and that obtained from the averaged 1D rough surfaces result (red solid line). Therefore, considering that generating 2D rough surfaces takes much more time than 1D, in the following, we adopt this averaging operation to deal with the 2D rough surfaces.

## 2.2 Effective medium model for rough surface

We assume the rough layer is composed of *n* effective layers, and the thickness of each layer is set as *t*. With this treatment, each layer can be considered as an inhomogeneous medium composed of SiC and vacuum. EMT seeks the average permittivity of the inhomogeneous medium by field averaging [42] and is therefore utilized to obtain the effective dielectric function of each layer in this work. There are several formulae in EMT, such as the Bruggeman approximation, Maxwell Garnett (MG) approach and Lorentz-Lorenz (LL) [43]. In this paper, the Bruggeman approximation is adopted since it treats both SiC and vacuum on an equal, self-consistent basis [44], which is rational in our model . Moreover, it has been used by previous researchers, such as Aspnes *et al*. [43, 44] and Liu *et al*. [45], to investigate the optical properties of samples with rough surfaces. Treating the SiC constituent in each layer as spherical particles embedded in the effective homogeneous medium, the Bruggeman approximation takes the form of [42]

$$f\frac{\varepsilon_{\text{SiC}} - \varepsilon_{\text{eff}}}{\varepsilon_{\text{SiC}} + 2\varepsilon_{\text{eff}}} + (1-f)\frac{\varepsilon_{\text{vac}} - \varepsilon_{\text{eff}}}{\varepsilon_{\text{vac}} + 2\varepsilon_{\text{eff}}} = 0 \quad (3)$$

where $\varepsilon_{\text{SiC}}$ and $\varepsilon_{\text{vac}}$ represent the dielectric function of SiC and vacuum respectively; $\varepsilon_{\text{eff}}$ and $f$ are the effective dielectric function and the filling fraction of SiC of the interested effective layer, respectively. The dielectric function of SiC is described by Lorentz model as [46]

$$\varepsilon(\omega) = \varepsilon_{\infty}\left(\frac{\omega^2 - \omega_{\text{LO}}^2 + i\Gamma\omega}{\omega^2 - \omega_{\text{TO}}^2 + i\Gamma\omega}\right) \quad (4)$$

where $\omega$ is the angular frequency, $\varepsilon_{\infty}$ is the high frequency dielectric constant, $\omega_{\text{LO}}$ is the longitudinal optical phonon frequency, $\omega_{\text{LO}}=1.825\times10^{14}$ rad/s, $\omega_{\text{TO}}$ is the transverse optical phonon frequency, $\omega_{\text{TO}} = 1.494\times10^{14}$ rad/s, $\Gamma$ is the damping factor, $\Gamma = 8.966\times10^{11}$ s$^{-1}$.

The effective dielectric function distribution of rough surfaces with different RMS height show different laws, which determine their respective characteristic of NFRHT. **Fig. 3** shows the real part of dielectric function varying with *dep* for $\sigma$ = 10 nm, 25 nm, 45 nm, from which the gradient distribution of dielectric function can be clearly observed. The value of dielectric function is proportional to the brightness of the region. It can be found that for the topmost several effective layers, the dielectric function is close to the value of vacuum, indicating that they have very slight effects on NFRHT. For a given $\omega$, the dielectric function approaches the value of bulk SiC as *dep* increases. By comparing the three figures, one can also find that the rough surface with larger $\sigma$, whose effective dielectric function gets the value of SiC at a larger *dep*, features a more gradual variation of dielectric function.

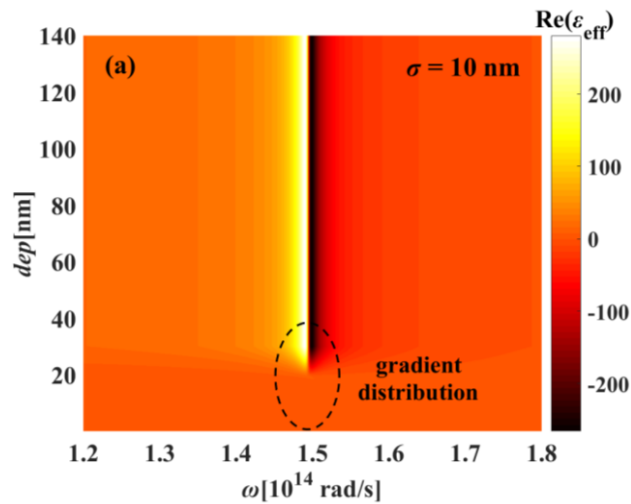

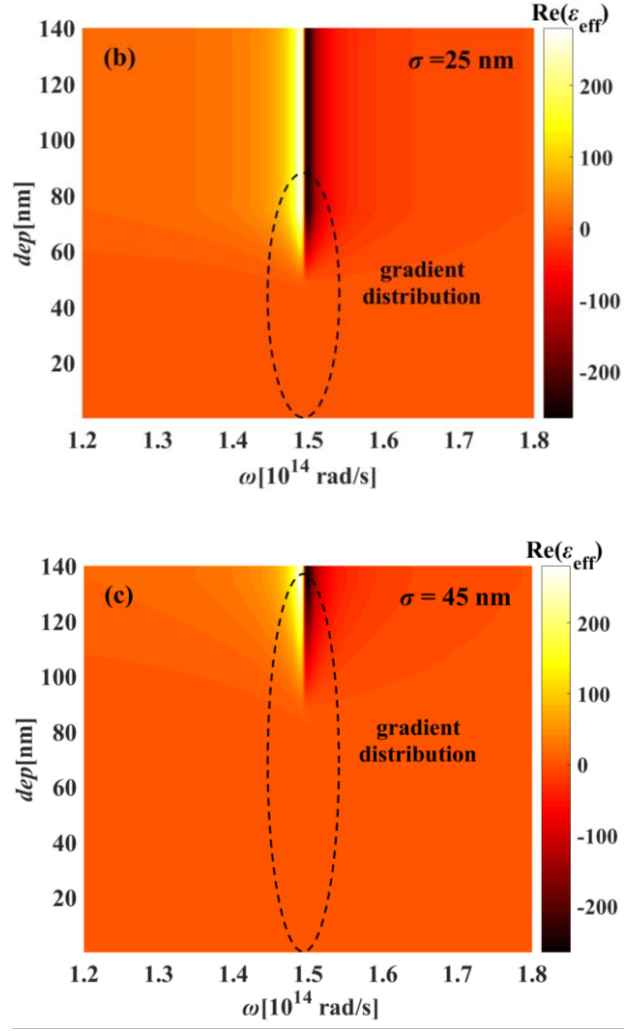

**Fig. 3** The variation of the real part of dielectric function with *dep* (the gradient distribution of dielectric function) for (a) $\sigma$ = 10 nm, (b) $\sigma$ = 25 nm, and (c) $\sigma$ = 45 nm. The value of dielectric function is proportional to the brightness of the region.

## 2.3 NFRHT between rough surfaces

By modeling the medium with rough surface as an effective multilayer medium, the NFRHT model between multilayer media can be used to calculate that between rough surfaces. As shown in **Fig. 4**, both media are divided into two parts: a multilayer system modeling the rough layer and a substrate. The temperatures of the emitter and receiver are $T_1$ and $T_2$, respectively ($T_1 > T_2$). To compare the NFRHT between rough surfaces to that between smooth plates, it is important to specify the definition of distance between rough surfaces since it has a great influence on NFRHT. We define the distance between the planes of $z = \sigma$ (marked by the blue dotted line in **Fig. 4**) of the two opposite rough surfaces as $d_{rms}$. According to the definition, it can be understood that $d_{rms}$ is an average distance between two opposite rough surfaces. Another definition is the distance between the highest points of two opposite rough surfaces, which is referred to as $d_{top}$. These definitions are shown schematically in

**Fig. 4**.

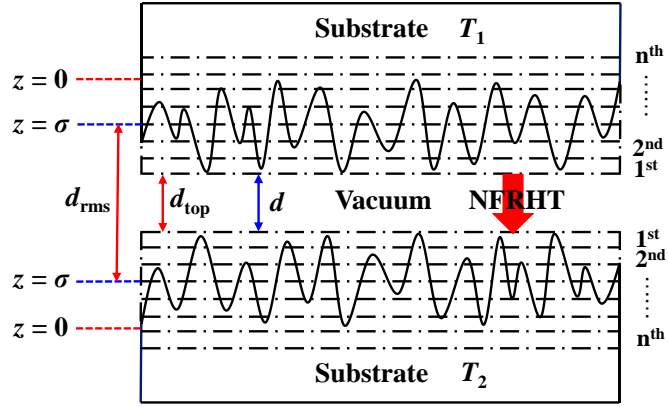

**Fig. 4** Schematic of the calculation model of NFRHT between random rough surfaces: an effective multilayer model. The red dotted line marks the location of the lowest point of the rough surface corresponding to $z = 0$ in **Fig. 2(a)** and the blue dotted line marks the location of RMS height corresponding to $z = \sigma$. $d_{rms}$ is defined as the distance between the blue dotted lines and $d_{top}$ is defined as the distance between the highest points of two opposite rough surfaces. The gap distance $d$ between two effective multilayer media is equal to $d_{top}$.

The heat flux of NFRHT between two multilayer media shown in **Fig. 4** is given by fluctuational electrodynamics [42]

$$q_{net} = \frac{1}{\pi^2} \int_0^\infty d\omega [\Theta(\omega, T_1) - \Theta(\omega, T_2)] \int_0^\infty s(\omega, \beta) d\beta \tag{5}$$

where $\Theta(\omega, T) = \hbar\omega / [\exp(\hbar\omega / k_B T) - 1]$ is the mean energy of a Planck oscillator at the frequency $\omega$ and the equilibrium temperature $T$, $\hbar$ is the Planck constant divided by $2\pi$, $k_B$ is the Boltzmann constant, $\beta$ is the component of the wavevector parallel to the interface. $s(\omega, \beta)$ is an exchange function, which equals the summation of the contribution of propagating waves ($\beta < \omega/c$) and evanescent waves ($\beta > \omega/c$), i.e., $s(\omega, \beta) = s_{prop}(\omega, \beta) + s_{even}(\omega, \beta)$. These two contributions are represented by $s_{prop}(\omega, \beta)$ and $s_{even}(\omega, \beta)$ respectively and are expressed as [1]

$$s_{prop}(\omega, \beta) = \frac{\beta(1-\rho_{01}^s)(1-\rho_{02}^s)}{4\left|1 - R_{01}^s R_{02}^s e^{i2\gamma_0 d}\right|^2} + \frac{\beta(1-\rho_{01}^p)(1-\rho_{02}^p)}{4\left|1 - R_{01}^p R_{02}^p e^{i2\gamma_0 d}\right|^2} \tag{6}$$

$$s_{even}(\omega, \beta) = \frac{\beta \operatorname{Im}(R_{01}^s) \operatorname{Im}(R_{02}^s) e^{-2\operatorname{Im}(\gamma_0)d}}{\left|1 - R_{01}^s R_{02}^s e^{-2\operatorname{Im}(\gamma_0)d}\right|^2} + \frac{\beta \operatorname{Im}(R_{01}^p) \operatorname{Im}(R_{02}^p) e^{-2\operatorname{Im}(\gamma_0)d}}{\left|1 - R_{01}^p R_{02}^p e^{-2\operatorname{Im}(\gamma_0)d}\right|^2} \tag{7}$$

In Eq. (6) and Eq. (7), $\gamma_0 = \sqrt{k_0^2 - \beta^2}$ is the component of the wavevector perpendicular to the interface, and $k_0 = \omega/c$ is the magnitude of the wavevector in vacuum. $R_{0i}^j$ is the effective reflection coefficient for $j$ ($j$ = s, p) polarization between vacuum and medium $i$ ($i$ = 1 for the upper plate and $i$ = 2 for the lower plate). In this paper, $R_{0i}^{j}$ is given by transfer matrix method (TMM). $\rho_{0i}^{j} = |R_{0i}^{j}|^2$ is the effective reflectivity. It is worth noting that $d$, the gap distance between two effective multilayer media, is equal to $d_{\text{top}}$ of the rough surfaces system when executing the calculations, as can be understood unambiguously through **Fig. 4**. The relation between $d_{\text{rms}}$ and $d_{\text{top}}$ is $d_{\text{rms}} = d_{\text{top}} + 2\times(3\sigma - \sigma)$. The heat flux calculated using $d = d_{\text{top}}$ (referred to as "the heat flux at $d_{\text{rms}}$") is compared with the heat flux between two smooth plates separated by $d_{\text{rms}}$ ($= d_{\text{top}} = d$ for smooth plates) to show the impact of roughness. Moreover, if the spectral heat flux needs to be calculated, one should not integrate over the angular frequency $\omega$ in Eq. (5), that is

$$q_\omega = \frac{1}{\pi^2}[\Theta(\omega,T_1) - \Theta(\omega, T_2)] \int_0^\infty [s_{\text{prop}}(\omega,\beta) + s_{\text{even}}(\omega,\beta)]d\beta \qquad (8)$$

## 2.4 Convergence verification and validation of the effective multilayer model

To model the medium with rough surface using the effective multilayer model, the rough layer is divided into multiple layers. Since the number of layers is not infinite, the obtained dielectric function distribution becomes dependent on the number of layers. With the increase of the number of effective layers, the effective multilayer model could better simulate the rough surface. When the number of layers increases to a certain value, the calculation result of the model will converge and can be considered reliable. Therefore, a convergence validation needs to be conducted.

For the rough surface with parameters in Section 2.1 (i.e., $N$ = 1000, $rL$ = 50 μm, $\tau$ = 0.1 μm, $\sigma$ = 10 nm), 4 cases are investigated: $n$ = 30, $t$ = 1 nm; $n$ = 60, $t$ = 0.5 nm; $n$ = 150, $t$ = 0.2 nm; $n$ = 200, $t$ = 0.15 nm. In all the cases, the total thickness of layers is equal to 30nm, which equals $3\sigma$. It is noted that the rough surface generated in each calculation is different because of the randomness of the algorithm used. In order to eliminate the influence of randomness, in each case 1000 rough surfaces are generated and the average value of $f$ is calculated. **Fig. 5** gives the calculation results of spectral heat flux for the four cases. The temperatures of emitter and receiver are set as 400 K and 300 K respectively and the gap distance is $d_{\text{rms}}$ = 200 nm. As shown in **Fig. 5**, when $n$ is relatively small, e.g., $n$ = 30 and $n$ = 60, the peak of the heat flux blue-shifts as $n$ increases, implying that the calculation results of our model are not reliable. However, this undesired phenomenon can be eliminated by increasing the number of layers. In the case of $n$ = 150, $t$ = 0.2 nm and $n$ = 200, $t$ = 0.15 nm, the spectral heat flux converges and no longer changes as the number of layers increases (by comparing the red line with rhombic markers pink line with trigonal markers in **Fig. 5**, verifying the reliability and stability of our mode in these cases. Thus, a convergent and accurate results can be obtained using $n \geq 150$. The convergence verification demonstrates that it is reasonable to take the number of effective layers as 150 when calculating the following cases.

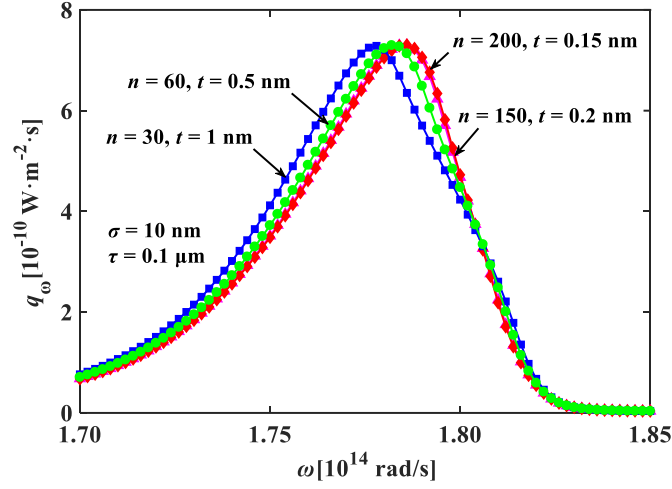

**Fig. 5** The calculation results of spectral heat flux for the four cases: $n = 30$, $t = 1$nm; $n = 60$, $t = 0.5$ nm; $n = 150$, $t = 0.2$ nm; $n = 200$, $t = 0.15$ nm. The parameters of rough surfaces are the same as that of **Fig. 2**(a): $\sigma = 10$ nm, $\tau = 0.1$ μm. When calculating the spectral heat flux, we set $T_1 = 400$ K, $T_2 = 300$ K, and $d_{rms} = 200$ nm.

Before using he effective multilayer model, it is necessary to validate it. Here, we employ the spectral heat flux indirectly given by Biehs and Graffet [34] using the second-order perturbation theory to verify the validation of the proposed model in both far and near field. In **Fig. 6**, we show the spectral heat flux calculated by the proposed method and that given by Ref. [34] at $d_{rms} = 5000$ nm (a) and $d_{rms} = 500$ nm (b). The parameters in our calculations are set as the same as those in Ref. [34]. As can be seen, the heat flux spectrum profile predicted by the proposed method is close to that given by perturbation theory, except for some slight differences near the peak locations. These differences may be caused by the difference between the exact profiles of rough surfaces generated in these two studies. One issue that needs to be pointed out is that the perturbation theory is valid when $\sigma$ is the smallest length involved in the problem, i.e., $\sigma \ll \min\{\tau, d, \lambda_{th}\}$ [47], where $\lambda_{th}$ is the characteristic wavelength of thermal radiation which equals 9.7 μm at room temperature. So in our following work, only the cases of relatively small $\sigma$ ($\sigma \ll \min\{\tau, d, \lambda_{th}\}$) are considered.

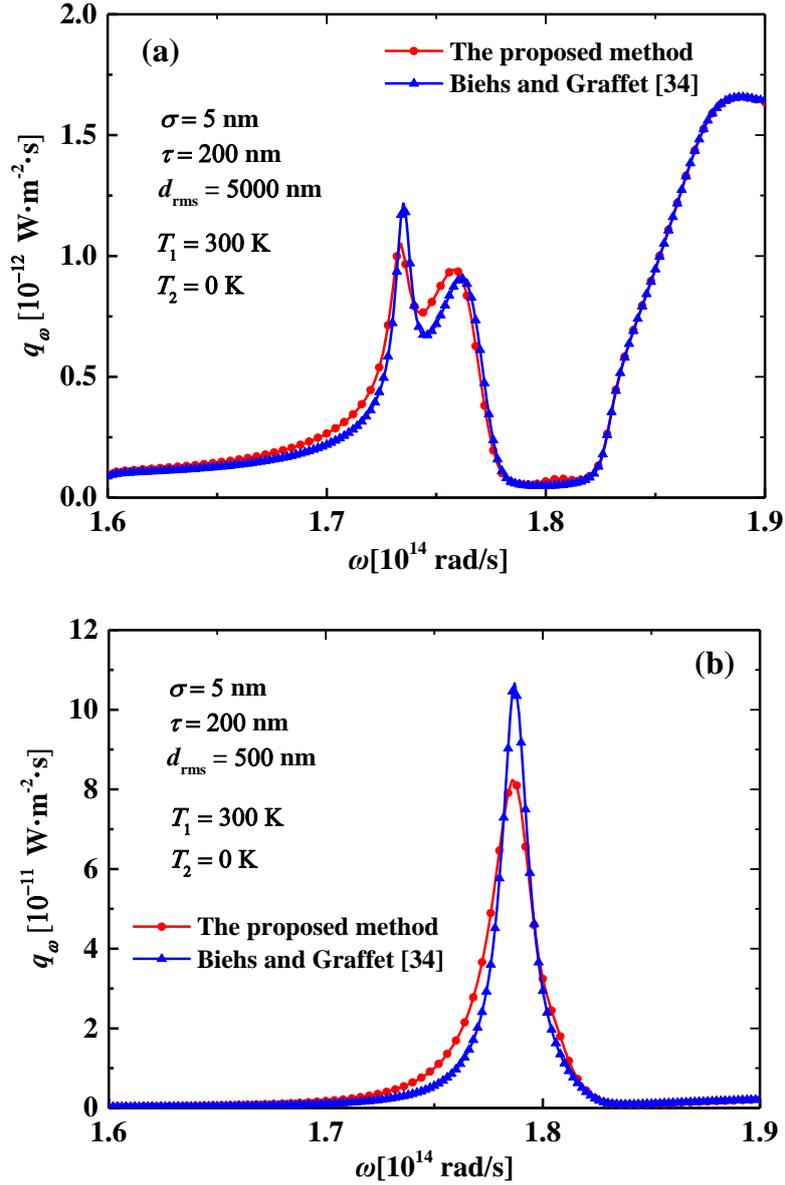

**Fig. 6** Comparison between the spectral heat flux calculated by the proposed method and that obtained from Ref. [34] for (a) $d_{rms}$ =5000 nm and (b) $d_{rms}$ =500 nm. Other parameters are shown in the figure.

## 3 Results and discussion

### 3.1 SPhPs resonance of the effective layers

SPhPs in different interfaces interact with each other, creating new dispersion relations different from that in any single interface [48]. The dispersion relation of surface polaritons (SPhPs for SiC) in single interface is written as [49]

$$\beta = \frac{\omega}{c}\sqrt{\frac{\varepsilon_1(\omega)\varepsilon_2(\omega)}{\varepsilon_1(\omega)+\varepsilon_2(\omega)}} \qquad (9)$$

where, $\beta$ is the component of wavevector parallel to the interface, $\varepsilon_1$ and $\varepsilon_2$ is the dielectric function for the two media on both sides of the interface, receptively. If one of the two media is vacuum (e.g., $\varepsilon_1 = 1$), Eq. (9) can be simplified to

$$\beta = \frac{\omega}{c}\sqrt{\frac{\varepsilon_2(\omega)}{1+\varepsilon_2(\omega)}} \qquad (10)$$

Eq. (10) indicates that when the equation $1 + \varepsilon_2(\omega) = 0$ is satisfied, the wavevector becomes extremely large, which implies the resonance of surface polariton. Then it can be deduced that the solution of equation $1 + \varepsilon_{\text{eff, i}}(\omega) = 0$ represents the resonance frequency of the SPhPs on the interface between the $i^{\text{th}}$ layer and vacuum.

**Fig. 7**(a) gives the dielectric function of effective medium with different SiC filling fraction. As can be seen, with the reduction of fraction, the dielectric function gets farmer from the value of SiC and approaches that of vacuum. The inset of **Fig. 7**(a) shows the intersections of the dielectric function curves of different filling fraction and the straight line $\text{Re}(\varepsilon) + 1 = 0$. These intersections represent the solutions of $1 + \varepsilon_{\text{eff}}(\omega) = 0$, with the abscissas of the intersections corresponding to the resonance frequencies of SPhPs. In **Fig. 7**(b), we give the resonance frequency, $\omega_{\text{res}}$, by solving the equation $1 + \varepsilon_{\text{eff}}(\omega) = 0$ for different filling fraction. As the filling fraction decreases, the resonance frequency is shifted away from the resonance frequency of SiC bulk, $\omega_{\text{res,SiC}} = 1.786\times10^{14}$ rad/s, getting closer to its lower limit $\omega_{\text{TO}}$, $1.494\times10^{14}$ rad/s. Biehs *et al*. [50] studied the surface modes of SiC plate with air porous inclusions modeled using the Maxwell-Garnett effective medium theory. Although different inclusion shape and equivalent method are used in that work, a similar redshift of SPhPs resonance was observed. Nevertheless, it is noted that when $f$ is smaller than 0.36, the effective dielectric function is larger than 0 in the whole frequency region, limiting $\beta$ within $\omega/c$ (due to Eq. (10)), the critical value of propagating waves ($\beta < \omega/c$) and evanescent waves ($\beta > \omega/c$). This indicates that when $f$ is very small, SPhPs cannot be excited on the interface between vacuum and the corresponding layer. For the effective layers whose $f$ are large enough to support SPhPs but smaller than unity, SPhPs with red-shifted resonance can be excited. (see **Fig. 7**(b)). These red-shifted SPhPs couple with each other and give birth to new surface modes that dominate the NFRHT.

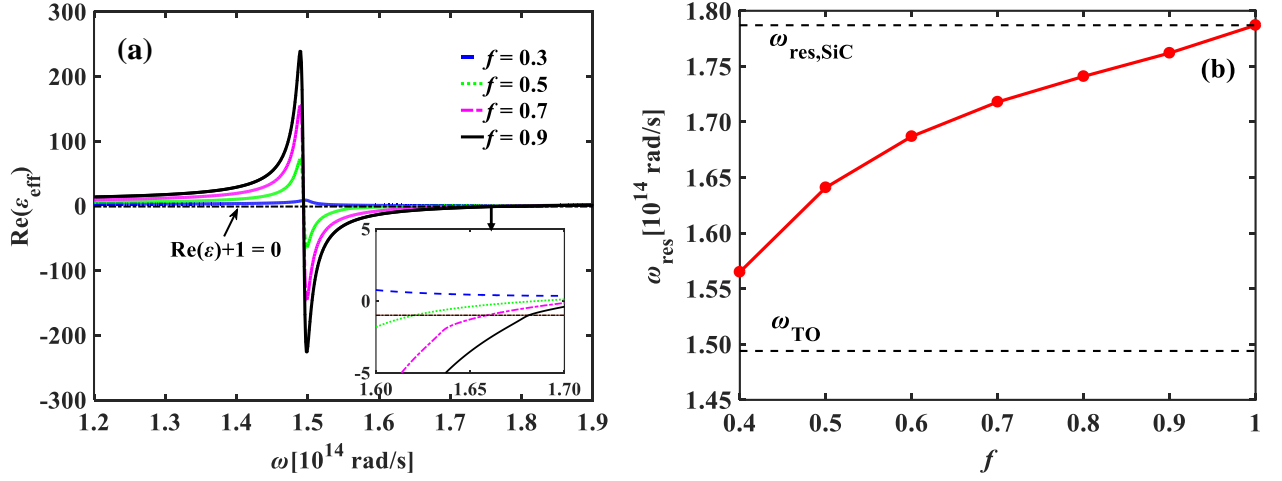

**Fig. 7** (a) Dielectric function of effective medium with different SiC filling fraction. $f$ = 0.3, 0.5, 0.7, 0.9. The black dotted line is the reference line Re($\varepsilon$)+1 = 0. The inset shows the intersections of the dielectric function curves and the straight line Re($\varepsilon$)+1 = 0. The abscissas of intersections correspond to the resonance frequencies of SPhPs. (b) The dependence of $\omega_\text{res}$ on filling fraction given by solving the equation $1+\varepsilon_\text{eff}(\omega) = 0$. The upper and lower limits are $\omega_\text{res,SiC}$ = 1.786×10$^{14}$ rad/s and $\omega_\text{TO}$ = 1.494×10$^{14}$ rad/s, respectively.

## 3.2 LDOS in rough surface/vacuum system

The local density of states (LDOS) is the number of electromagnetic modes per unit frequency at per unit volume. It is a fundamental quantity and can provide a quantitative understanding of the enhanced radiative heat transfer [1]. The monochromatic LDOS in a certain location above an interface between vacuum and a multilayer (or a bulk) medium $i$ is derived as [51]

$$D(h,\omega) = \frac{\omega^2}{2\pi^2 c^3}\left\{\int_0^1 \frac{\kappa d\kappa}{p}\left\{2+\kappa^2[\text{Re}(R_{0i}^s e^{2i p\omega h/c})+\text{Re}(R_{0i}^p e^{2i p\omega h/c})]\right\} \right.$$
$$\left. +\int_1^\infty \frac{\kappa^3 d\kappa}{|p|}\left[\text{Im}(R_{0i}^s)+\text{Im}(R_{0i}^p)\right]e^{-2|p|\omega h/c}\right\} \quad (11)$$

where, $\kappa = \beta c/\omega$, when $\kappa \leq 1$, $p = \sqrt{1-\kappa^2}$; when $\kappa > 1$, $p = i\sqrt{\kappa^2-1}$. $R_{0i}^j$ is the same as that mentioned above and is obtained through transfer matrix method (TMM). $h$ represents the distance between a point in vacuum and the interface. Larger LDOS stands for more electromagnetic modes at that frequency and location, providing more channels for energy transfer. For rough surfaces, the $z$-coordinates of the points on the interface are different. It is necessary to define an average height to describe the location of a given point in vacuum. Here, referring to the definition of $d_\text{rms}$, we define $h_\text{t}$, an average height, as the distance from a given point in vacuum to the plane where $z = \sigma$ in the medium. (shown in **Fig. 8**). Then one can calculate the LDOS above a rough surface using Eq. (11) by replacing

$h$ with $h_t$.

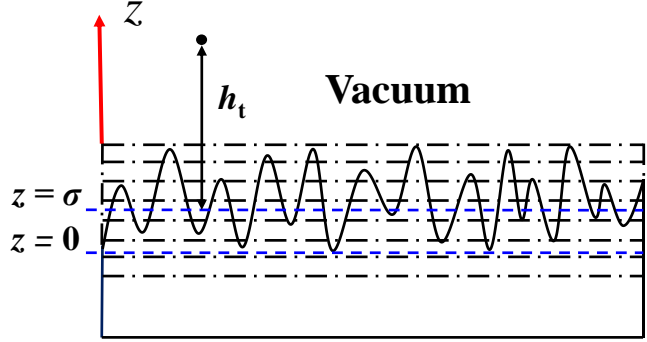

**Fig. 8** Schematic of the calculation model of LDOS in vacuum/rough surface system and the definition of $h_t$. $h_t$ is defined as the distance from a given point in vacuum to the plane where $z = \sigma$ in the medium.

Here, we inspect two situations: $h_t = 100$ nm (**Fig. 9**(a)) and $h_t = 500$ nm (**Fig. 9**(b)). For each situation, 3 RMS heights ($\sigma = 10$ nm, $\sigma = 25$ nm, $\sigma = 45$ nm) are investigated and the LDOS at $h = h_t$ above the interface between vacuum and a smooth plate is also shown for comparison. $\tau$ is set to be 0.1 μm as the controlled variable. For $h_t = 100$ nm (shown in **Fig. 9**(a)), the vertical axis, which represents the value of LDOS, is scaled logarithmically. As $\sigma$ increases, the LDOS in the frequency region $\omega < \omega_{\text{res,SiC}}$ increases significantly while that in the frequency region $\omega > \omega_{\text{res,SiC}}$ increases slightly. Meanwhile, the peak of LDOS is broadened and shifts to the lower frequency. In the case of $h_t = 45$ nm, the peak comes to $1.750 \times 10^{14}$ rad/s with a value of $3.20 \times 10^6$, 6 times the LDOS above the smooth surface. This is the result of the mutual coupling of SPhPs among the effective layers of the rough surface. The coupling of the SPhPs, which supported by the effective layers with small $f$, provides more electromagnetic states with lower frequency (see **Fig. 7**(b)), causing larger LDOS in the corresponding frequency region. Biehs and Greffet [47] also investigated the LDOS above a rough surface using perturbation theory used in [34]. A similar broadened and red-shifted dispersion relation was reported in that work. This consistency once more validates our method. More interestingly, that work attribute the phenomenon of broadening and red shift to the scattering of surface phonons into other states induced by the roughness, which is in agreement with our discussion about SPhPs coupling above.

However, the case of $h_t = 500$ nm shows entirely different laws. As can been seen in **Fig. 9**(b), the peak of LDOS decreases dramatically as $\sigma$ increases with its location almost stays at $\omega_{\text{res,SiC}}$. One can understand it considering that the effect of roughness is weak when a location far from the surface is investigated. In fact, either the SPhPs supported by the bulk SiC or those supported by the effective layers with small $f$ decay exponentially with the distance from the surface. By comparing **Fig. 9**(a) and (b), it can be inferred that the SPhPs supported by the effective layers with small $f$ decay more drastically. As a result, the states with lower frequencies provided by these effective layers are submerged by the states provided by the SiC bulk, which peaks at $\omega_{res,SiC}$. This explains why the peak of LDOS at $h_t$ = 500 nm cannot be red-shifted obviously as the case $h_t$ = 100 nm does. In summary, the effect of roughness is significant only when the distance is not too large. This conclusion offers us a reference for the selection of distance in what follows. On the other hand, as $\sigma$ increases, the SiC bulks on both sides of the vacuum gap are brought father from each other, which decreases the peak of LDOS as shown in **Fig. 9**(b).

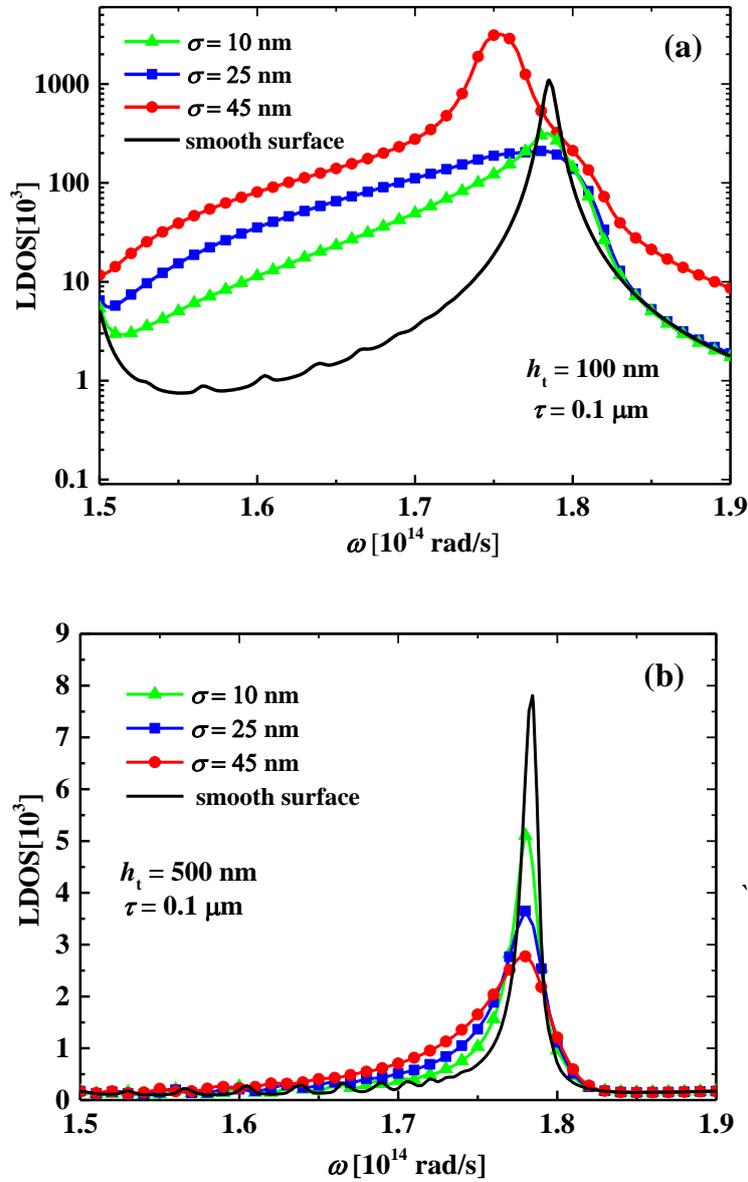

**Fig. 9** The LDOS in a vacuum/rough surface system for (a) $h_t$ = 100 nm, (b) $h_t$ = 500 nm. In each figure, the LDOS of the case of $\sigma$ = 10 nm (green line with triangle markers), 25 nm (blue line with square markers), 45 nm (red line

with circle markers) are shown. $\tau$ is set to be 0.1 μm as the controlled variable. The LDOS at $h = h_t$ above the interface of a smooth surface is also shown with black line for comparison.

## 3.3 The effect of RMS height on NFRHT

Based on LDOS, the NFRHT between rough surfaces is studied and compared with that between smooth plates. The distance $d_{rms}$ of rough surfaces system is adopted and set as 200 nm (for smooth surfaces, $d$ = 200 nm). Here, we choose 200 nm to ensure that $\sigma$ has a significant effect on heat transfer and the two opposite rough surfaces do not touch each other. The temperatures of the emitter and receiver are 400 K and 300 K respectively and $\tau$ is set to be 0.1 μm as the controlled variable. The rough surfaces of $\sigma$ = 5 nm, $\sigma$ = 20 nm, $\sigma$ = 35 nm, $\sigma$ = 45 nm are investigated, as well as the smooth plates ($\sigma$ = 0).

**Fig. 10**(a) shows the spectral heat flux between rough surfaces with above RMS heights. We set $\sigma$ of the rough surfaces on both sides of the vacuum gap to be the same to clarify its effect on NFRHT. As shown in the figure, when $\sigma$ = 5 nm, the effect of surface roughness is not significant and the heat flux still peaks at $\omega_{res,SiC}$ = 1.786×10$^{14}$ rad/s, the same as that between smooth plates. This result once more verifies the reliability of our effective model for rough surfaces. However, the maximum value of heat flux is not as large as that between smooth plates. As $\sigma$ increases, the peak of heat flux is red-shifted gradually. This red shift phenomenon is similar to that reported by Chen and Xuan in Ref. [37], which is about Drude material. In what follows, we refer to the new peak of spectral heat flux as $\omega_{pea}$. In **Fig. 10**(b), we show the variation of $\omega_{pea}$ with $\sigma$. $\omega_{pea}$ decreases continuously with increasing $\sigma$ then remains as a constant when $\sigma$ is larger than a certain value. When $\sigma$ = 45 nm, it arrives at 1.750×10$^{14}$ rad/s, which is consistent with the peak of LDOS. In addition to the red shift of $\omega_{pea}$, the spectral heat flux increases in the lower frequency region ($\omega < \omega_{pea}$) with the increase of $\sigma$. In general, the heat flux spectrum is broadened with the peak red-shifted as $\sigma$ increases. Interestingly, Wang *et al*. [52] investigated the influence of parameters in Drude model on NFRHT between two smooth plates, and showed the broadening of the resonance heat flux spectrum and red shift of the resonance peak with increased scattering rate, which is similar to the above observation, indicating similar mechanism. That is, the change of parameter of the rough surface act equivalently as the tuning of of dielectic function.

The mechanism of the heat flux spectrum can be understood via the analysis of SPhPs resonance in Section 3.1 as follows. With $\sigma$ increasing, the effective layers with small *f* on both sides of the vacuum gap are brought more closer to each other, so the coupling of their SPhPs, whose resonances are red-shifted, plays a more important role in NFRHT. Meanwhile, as *dep* increases, although *f* gets larger, the surface waves are attenuated severely when it arrives at the vacuum gap due to losses of media, resulting in less influences of the interior effective layers and SiC bulk on NFRHT. The above two factors make the heat flux peak deviate from $\omega_{res,SiC}$ and reach $\omega_{pea}$. To get insight to the overall effect of roughness on NFRHT, the dimensionless total net heat flux (divided by that between SiC rough surfaces) which is determined by Eq.(5) is calculated and also shown in **Fig. 10**(b) as a function of $\sigma$. As can be seen, the total net radiative heat flux between rough surfaces exceeds the SiC smooth plates benchmark and gets larger with increasing $\sigma$. When $\sigma$ = 45 nm, the total net heat flux reaches 2.7 times that between smooth plates. This result holds promise for the improvement of energy utilization efficiency of devices with rough surfaces.

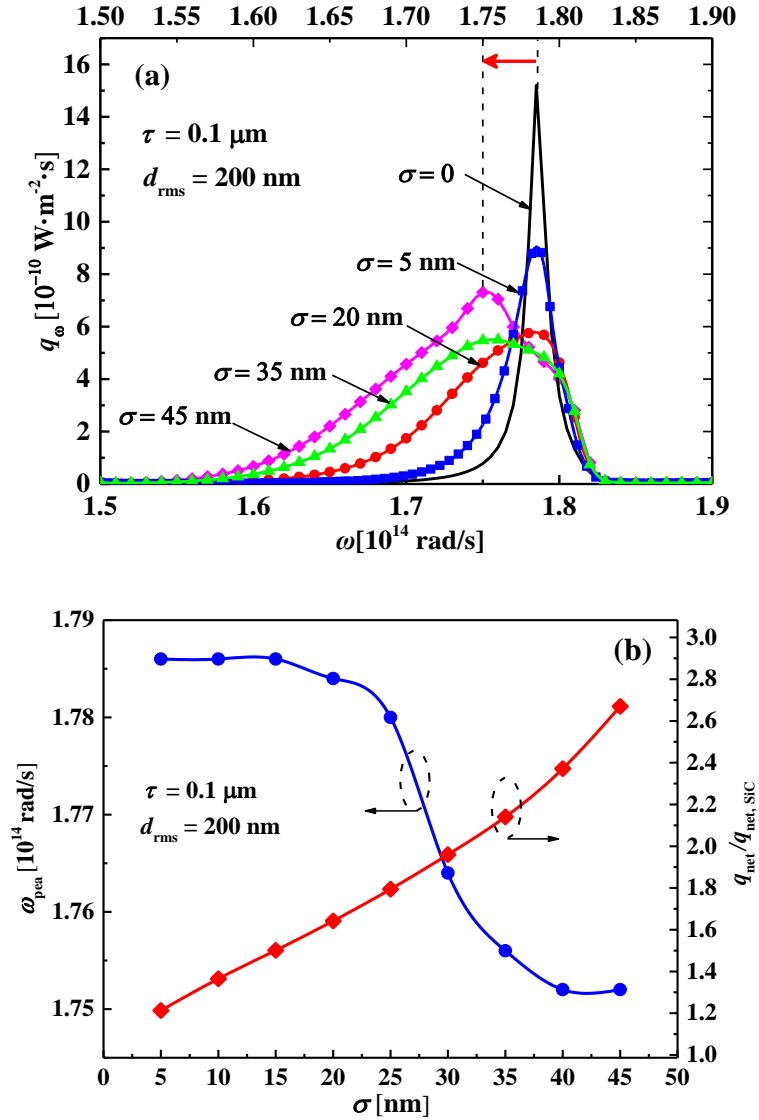

**Fig. 10** (a)Spectral heat flux between rough surfaces. The RMS heights($\sigma$) on both sides of the vacuum gap are set to be the same and the case of $\sigma$ = 5nm, 20 nm, 35 nm, 45 nm are shown in different colors and markers. The spectral heat flux between smooth plates is also shown for comparison.    (b)Blue line with circle markers corresponding to the left vertical axis: the angular frequency where the heat flux peaks, $\omega_{pea}$. Red line with rhombus markers corresponding to the right vertical axis: the dimensionless tot net heat flux ($q_{net}/q_{net,SiC}$). Both lines are functions of $\sigma$ which corresponds to the horizontal axis.

In order to uncover the physical mechanism of the NFRHT between rough surfaces more thoroughly, in this paper, we introduce the energy transmission coefficient (or photon tunneling probability). For nonmagnetic media, SPhPs could only be excited by electromagnetic waves in p-polarized states (TM waves) [49], so the contribution of electromagnetic waves in s-polarized states (TE waves) can be ignored. Meanwhile, when studying the NFRHT, the contribution of propagating waves is usually neglected as the magnitudes of their wavevectors are so small compared to that of evanescent waves. Therefore, what should be focused on is the energy transmission coefficient of evanescent waves in p-polarized state, which is written as [31]

$$\xi_p(\omega,\beta) = \frac{4\,\mathrm{Im}(R_{01}^p)\,\mathrm{Im}(R_{02}^p)e^{-2\mathrm{Im}(\gamma_0)d}}{\left|1-R_{01}^p R_{02}^p e^{2i\gamma_0 d}\right|^2} \tag{12}$$

From Eq. (12), it can be derived that $\xi_p(\omega,\beta)$ will be divergent when the relation $1-R_{01}^p R_{02}^p e^{2i\gamma_0 d}=0$ is satisfied, indicating that the contribution of SPhPs to NFRHT is infinity. The relation $1-R_{01}^p R_{02}^p e^{2i\gamma_0 d}=0$ is thus regarded as the dispersion relation of coupled SPhPs of multilayer system. But in actual media, due to the existence of loss, $1-R_{01}^p R_{02}^p e^{2i\gamma_0 d}=0$ could not be perfectly fulfilled, limiting $\xi_p(\omega,\beta)$ within unity, which will be achieved when every photon emitted by one substance with any direction can be totally absorbed by the other body [10].

In **Fig. 11**, the contour of energy transmission coefficient is plotted in the $\beta$-$\omega$ plane. The lateral wavevector $\beta$ is normalized by the wavevector in vacuum $k_0$ (= $\omega/c$). We can see a bright region with high value of energy transmission coefficient in the contour. The brighter the region, the higher the energy transmission coefficient of the region, indicating that the photons with such $\omega$ and $\beta$ are more possible to tunnel the vacuum gap, hence contributing to NFRHT. For smooth plates, as shown in **Fig. 11**(a), the two branches of bright regions represent the odd mode and the even mode of SPhPs, respectively, which converge with a large $\beta$ (as large as $40k_0$) at $\omega_{res,SiC}$. As $\beta$ increases, both branches of regions with high $\xi_p$ become parallel to the $\beta$-axis, implying that there are huge amounts of possible states with different $\beta$ for the corresponding $\omega$, thus providing more channels for NFRHT, in other words, increasing the heat flux. When $\sigma$ = 10 nm (shown in **Fig. 11**(b)), a high $\xi_p$ region spanning a large $\beta$ range ($10k_0 \sim 20k_0$) can be observed, which is associated with the peak of spectral heat flux. The reduction of $\beta$ (compare **Fig. 11**(b) with **Fig. 11**(a)) explains the decrease of heat flux compared to that between smooth plates, as shown in **Fig. 10**(a). Below the peak frequency, especially in $1.60\times10^{14}$ rad/s~$1.78\times10^{14}$ rad/s, the high $\xi_p$ region is broadened due to the coupling of SPhPs supported by the effective layers with small $f$, which increases the heat flux. As $\sigma$ increases further, the rough layer continues to enlarge high $\xi_p$ region and provides more coupled SPhPs modes with larger $\beta$ in $1.60\times10^{14}$ rad/s~$1.78\times10^{14}$ rad/s. In the cases of $\sigma$ =25 nm and $\sigma$ = 45 nm, the regions with non-zero $\xi_p$ attains about $25k_0$ and $30k_0$ at $\omega_{pea}$ = $1.780\times10^{14}$ rad/s and $1.750\times10^{14}$ rad/s, respectively. In a word, the coupled SPhPs modes in the rough layer enhance the NFRHT in lower frequency region and red-shift the heat flux peak, as described in **Fig. 10**.

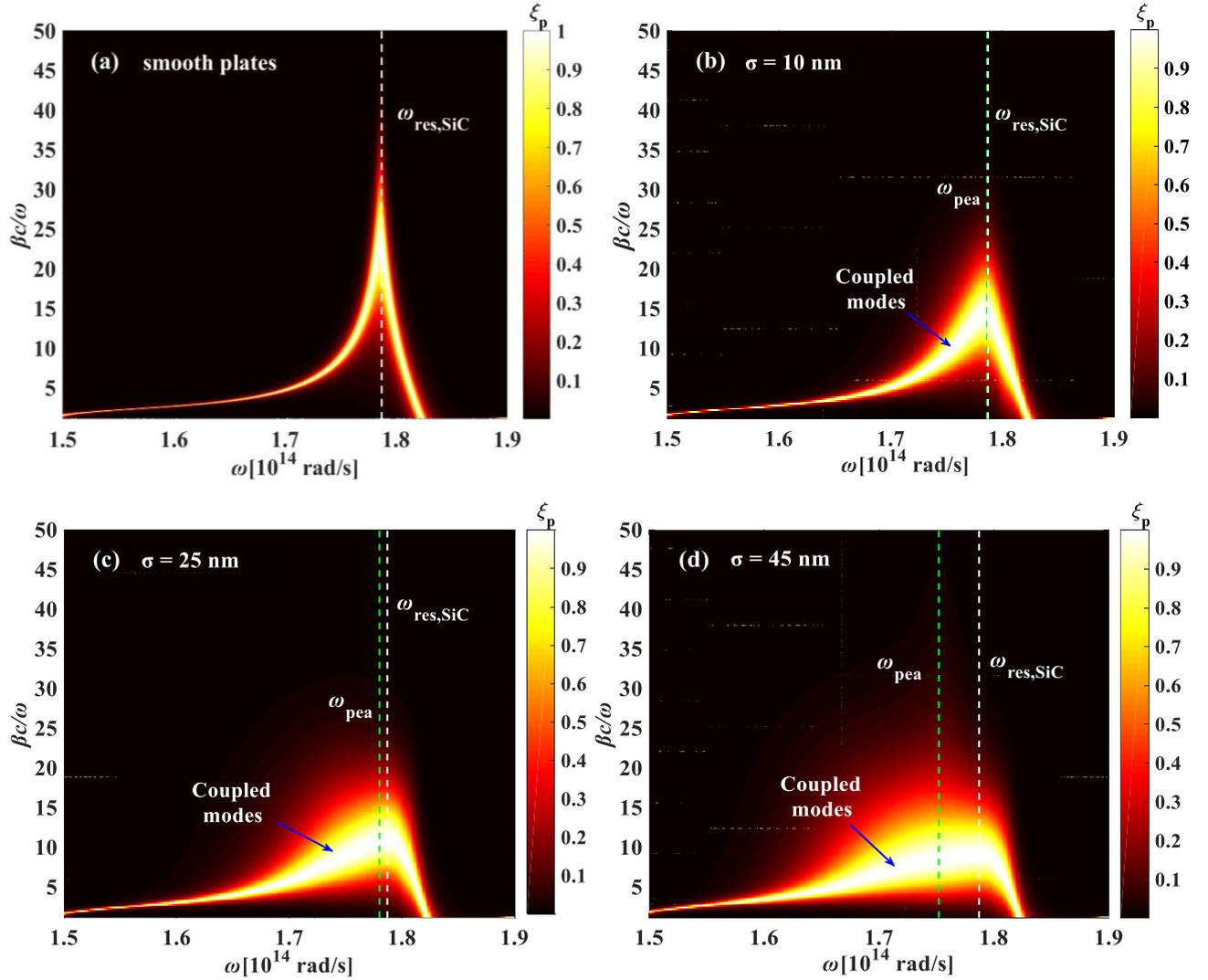

**Fig. 11** The contour of energy transmission coefficient $\xi_p$ in the $\beta$-$\omega$ plane for (a) smooth plates (b) $\sigma$ = 10 nm (c) $\sigma$ = 25 nm and (d) $\sigma$ = 45 nm. The wavevector $\beta$ is normalized by the wavevector in vacuum $k_0$ (= $\omega/c$). $\tau$ = 0.1 μm, $d_{rms}$ = 200 nm for all the cases. The white dashed lines represent $\omega_{res,SiC}$, while the green dashed lines in the four figures represent $\omega_{pea}$ for the corresponding $\sigma$. A red shift of the high $\xi_p$ region with large wavevector can be observed by comparing the four figures.

## 3.4 Feasibility of analyzing the effect of correlation length

As has been discussed in Section 2.1, another parameter describing the geometric characteristic of random rough surfaces is the correlation length, $\tau$. In this section, we will test the feasibility of our model when analyzing the effect of correlation length. Three cases are inspected here: $\tau = 0.05$ μm, $\tau = 0.1$ μm, and $\tau = 0.5$ μm. For all cases, $\sigma$ is set to be 25 nm as the controlled variable. In **Fig. 12**, the spectral heat flux is calculated for the three cases above. ($T_1 = 400$ K, $T_2 = 300$ K, $d_{rms} = 200$ nm). Obviously, the spectral heat flux hardly does not vary with $\tau$, which indicates that the correlation length has no impact on NFRHT. That is not reliable considering that Chen and Xuan [38] have demonstrated the influence of correlation on NFRHT. Moreover, it can be easily predicted that when $\tau$ gets larger, the characteristic of rough surface would approach that of its smooth plate counterpart. In fact, as can be seen and understood from Eq. (1), once $\sigma$ is fixed, the probability that the random points appear on a given layer is fixed, regardless of the change of $\tau$. As a result, the filling fraction $f$ which is extracted from corresponding positions of Gaussian distributed random surface is independent on $\tau$. The same is true for the dielectric function distribution, leading to the inability of our model to predict the effect of $\tau$ on NFRHT. Indeed, it is not an easy subject to include the correlation length in an effective medium model [45], which remains to be further studied.

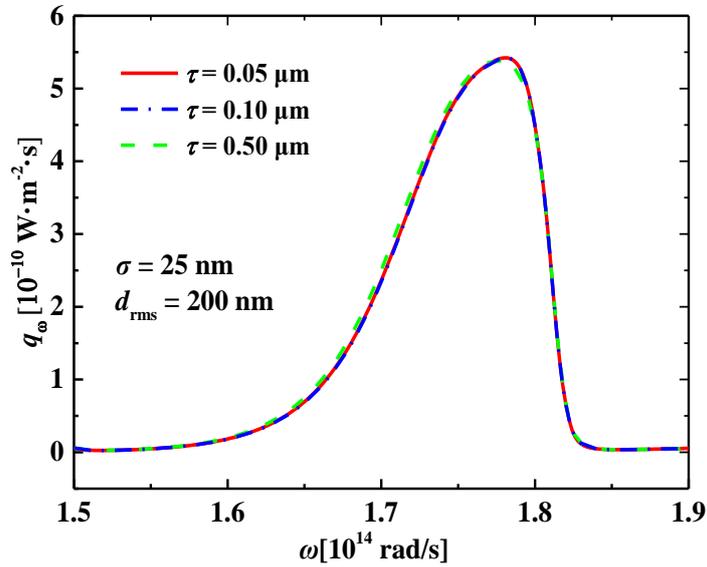

**Fig. 12** The spectral heat flux between rough surfaces with different correlation length $\tau$ ($\tau = 0.05$ μm, 0.1 μm, 0.5 μm). $\sigma = 30$ nm for all the three cases. The gap distance is set as $d_{rms} = 200$ nm. The temperatures of emitter and receiver are set as 400 K and 300 K, respectively.

## 4 Conclusions

The NFRHT between SiC rough surfaces is studied by modeling the rough layer as an effective multilayer medium with gradient distribution of dielectric function. The effective layers with small filling fraction of SiC feature lower SPhP resonance frequencies than bulk SiC. The coupling of SPhPs from the gradient distribution of dielectric function gives birth to new modes dominating the NFRHT. The peak of LDOS is red-shifted when roughness is considered for the location near the surface. As $\sigma$ increases, the peak of spectral heat flux is red-shifted while the NFRHT below the peak frequency is enhanced. The underlying mechanism is due to the coupling of SPhPs on the interfaces inside the rough layer. The total net radiative heat flux between rough surfaces exceeds that between smooth plates and increases with the increase of $\sigma$ when $d_{rms}$ is fixed. Nevertheless, this model fails to describe the effect of correlation length on NFRHT due to the losing of structure details in the effective multilayer model. In general, this work provides a new approach for the simulation and understanding of the NFRHT between rough surfaces, which may be helpful for related applications using rough surfaces in energy conversion and management.

## 5 Acknowledgments

The support by the National Natural Science Foundation of China (Grant nos. 51336002) and the Fundamental Research Funds for the Central Universities (Grant No. HIT.BRETIII.201415) was gratefully acknowledged. ZMZ wishes to thank the sponsorship of the Basic Energy Sciences, U.S. Department of Energy (DE-SC0018369).